\pdfoutput=1
\documentclass[12pt]{article}

\font\grande=cmr9.5 scaled \magstep4
\font\medio=cmr9.5 scaled \magstep2
\outer\def\beginsection#1\par{\medbreak\bigskip
      \message{#1}\leftline{\bf#1}\nobreak\medskip
\vskip-\parskip
      \noindent}
\usepackage{graphicx} 
\begin{document}
\bibliographystyle {unsrt}

\titlepage
\begin{flushright}
CERN-TH-2018-215
\end{flushright}
\vspace*{1.5cm}

\begin{center}
{\grande Polarized backgrounds of relic gravitons}\\
\vspace{6mm}
\vspace{15mm}     
 Massimo Giovannini 
 \footnote{Electronic address: massimo.giovannini@cern.ch} \\
\vspace{1cm}
{{\sl INFN, Section of Milan-Bicocca, 20126 Milan, Italy}}

\vspace*{1cm}
\end{center}

\centerline{\medio  Abstract}
\vspace{5mm}
The  polarizations of the tensor modes of the geometry evolving in cosmological 
backgrounds are treated as the components of a bispinor whose dynamics follows 
from an appropriate gauge-invariant action.  This novel framework bears a close analogy 
with the (optical) Jones calculus and leads to a compact classification of the various interactions 
able to polarize the relic gravitons.
\vskip 0.5cm

\nonumber
\noindent

\vspace{5mm}

\vfill
\newpage
The stochastic backgrounds of gravitational radiation may be formed by relic gravitons
parametrically amplified in the early Universe, as suggested long ago by Grishchuk \cite{gr}. 
In a general relativistic context the action of the relic gravitons has been derived, for the first time, by Ford and Parker 
\cite{fordp}. In conventional inflationary models the two polarizations of the gravitational waves do not 
interact \cite{star1} and consequently the low-frequency branch of the spectrum, ranging between 
the aHz and $100$ aHz, is unpolarized \cite{ir1}. The same conclusion holds at higher frequencies, 
e.g. in the mHz band and in the audio band. Hereunder we shall conventionally refer to the audio band 
as the region between few Hz and $10$ kHz;  standard prefixes will be used throughout when needed (e.g. $1\,\mathrm{mHz}= 10^{-3} \,\, \mathrm{Hz}$). 
 
At late times the evolution of the relic gravitons is affected by various anisotropic stresses whose transverse and traceless modes could induce a certain degree of polarization. After neutrino decoupling the corresponding anisotropic stress slightly suppresses the relic graviton background \cite{anis1} but it is unable to polarize the spectrum either in the audio band or in the mHz range. The polarization of the graviton background induced by  the anisotropic stresses typically involves a limited interval of frequencies  reflecting the physical properties of the source. For instance the anisotropic stress of the hypermagnetic knots (i.e. maximally gyrotropic configurations of the hypermagnetic fields) could polarize the stochastic backgrounds of relic gravitons over intermediate frequencies approximately ranging between few $\mu$Hz and $10$ kHz \cite{anis3}. According to a complementary perspective the mutual interaction of the 
two tensor polarizations might be ultimately responsible for the overall polarization of the cosmic graviton background. The purpose of this paper is then to scrutinize and classify the mutual interactions of the tensor polarizations in cosmological backgrounds by expressing their gauge-invariant action in terms of appropriate bispinors whose components coincide with the polarized amplitudes.

The tensor fluctuations of conformally flat background geometries are defined as 
$g_{\mu\nu}(\vec{x},\tau) = \overline{g}_{\mu\nu} + 
\delta^{(1)}_{t} g_{\mu\nu}$  where $\overline{g}_{\mu\nu}= a^2(\tau) \eta_{\mu\nu}$, $a(\tau)$ 
is the scale factor and $\eta_{\mu\nu}$ is the Minkowski metric. Overall the metric fluctuations 
contain ten independent components; only two describe the cosmic gravitons
and are parametrized in terms of a rank-two tensor in three spatial dimensions, i.e.   
$\delta^{(1)}_{t} g_{ij} = - a^2(\tau) h_{ij}(\vec{x},\tau)$ 
where $h_{ij}$ is divergenceless and traceless (i.e. $h_{i}^{\,\,\,i} = \partial_{i}\,h^{\,\,\,i}_{j}=0$). 
Since the tensor modes of the geometry are real quantities 
they can be  expressed in Fourier space as\footnote{Within the notations employed in this 
paper we have that $\ell_{P} = \sqrt{8 \pi G} = 1/\overline{M}_{P}$; $\overline{M}_{P}$ is the 
reduced Planck mass related to $M_{P}= 1/\sqrt{G}$ as $\overline{M}_{P} = M_{P}/\sqrt{8 \pi}$; $\tau$ will denote throughout the conformal time coordinate. We finally remind that in this paper
the Latin indices are all Euclidian and when this is the case there is no difference between covariant and contravariant components. }: 
\begin{equation}
h_{ij}(\vec{x},\tau) = \frac{\sqrt{2} \ell_{P}}{(2\pi)^{3/2}} \int \,\, d^{3} k \,\, h_{i j}(\vec{k},\tau) \,\, 
e^{ - i\, \vec{k}\cdot\vec{x}}, \qquad 
h_{ij}(\vec{k},\tau) = e_{ij}^{\otimes}\,\, h_{\otimes}(\vec{k}, \tau)  
+ e_{ij}^{\oplus}\,\, h_{\oplus}(\vec{k},\tau), 
\label{one}
\end{equation} 
where $h_{ij}^{*}(\vec{k},\tau) = h_{ij}(- \vec{k},\tau)$ and the factor $\sqrt{2} \ell_{P}$ appearing in Eq. (\ref{one}) is determined, as we shall see in a moment, by demanding that the action for each of the two polarizations canonically normalized. The two (orthogonal) polarizations are  $e_{ij}^{\otimes} = (\hat{m}_{i} \hat{n}_{j} + \hat{n}_{i} \hat{m}_{j})$ and $e_{ij}^{\oplus} = ( \hat{m}_{i} \hat{m}_{j} - \hat{n}_{i} \hat{n}_{j})$ where $\hat{m}$, $\hat{n}$ and $\hat{k}$ form a triplet of mutually orthogonal unit vectors in the three spatial dimensions (i.e. $\hat{m}\times\hat{n} = \hat{k}$). 

Following the spirit (if not the letter) of the Jones calculus\footnote{In optical applications the Jones calculus stipulates that the electric fields of the waves are organized in a two-dimensional column vector. In the analyses of optical phenomena the Jones approach is customarily contrasted with the Mueller calculus where the polarization is described by a four-dimensional (Mueller) column vector whose components are the four Stokes parameters \cite{robson}. } \cite{robson} the tensor polarizations 
can be  arranged in a bispinor, be it $\Psi$, whose components are given by $h_{\oplus}(\vec{k}, \tau)$ and by $h_{\otimes}(\vec{k}, \tau)$ respectively. The  action describing the dynamics of $\Psi$ must be invariant under infinitesimal diffeomorphisms,  it must contain (at most) two derivatives with respect to $\tau$ and it should reduce to the Ford-Parker action \cite{fordp} when the interactions between the polarizations are absent. Putting together these three requirements\footnote{It should be stressed that these requirements are physically complementary but conceptually separate.} we are led to the following expression: 
\begin{eqnarray}
S_{pol} &=& \frac{1}{2} \int d^{3} k \int d\tau \biggl\{ a^2(\tau)\biggl[ \partial_{\tau} \Psi^{\dagger} \partial_{\tau} \Psi - 
k^2 \Psi^{\dagger} \Psi  \biggr] +  \Psi^{\dagger} (\vec{v}\cdot\vec{\sigma}) \Psi +   \partial_{\tau} \Psi^{\dagger} (\vec{r}\cdot\vec{\sigma}) \partial_{\tau} \Psi
\nonumber\\
&+&  \Psi^{\dagger} (\vec{p}\cdot\vec{\sigma}) \partial_{\tau} \Psi +  \partial_{\tau} \Psi^{\dagger} (\vec{q}\cdot\vec{\sigma}) 
\Psi  \biggr\},\qquad \Psi = \left(\matrix{h_{\oplus}\cr h_{\otimes}} \right),
\label{twoa}
\end{eqnarray}
where $\vec{\sigma} =(\sigma_{1}, \, \sigma_{2},\, \sigma_{3})$ and $\sigma_{i}$ (with $i=1,\,\, 2,\,, 3$) are the three Pauli matrices.  The dagger denotes, as usual, the transposed and complex conjugate of the corresponding spinor or matrix. 
In the parametrization of Eq. (\ref{twoa}) the vector $\vec{r}(k,\tau)$ is dimensionless while the two vectors $\vec{v}(k,\tau)$, $\vec{p}(k,\tau)$, $\vec{q}(k,\tau)$ are all dimensional and may otherwise contain an arbitrary dependence on $k$. Finally, since the quantum Hamiltonian associated with the action (\ref{two}) must be Hermitian we are led to demand that $\vec{p} = \vec{q}$.  Thanks to this plausible requirement the terms containing a single conformal time derivative can be eliminated  by dropping a total time derivative while $\vec{b}$ is redefined as $\vec{v} \to \vec{b}= \vec{v} - \partial_{\tau} \vec{p}$. Therefore the canonical form of Eq. (\ref{two}) is always expressible as:
\begin{equation}
S_{pol} = \frac{1}{2} \int d^{3} k \int d\tau \biggl\{ a^2(\tau)\biggl[\partial_{\tau} \Psi^{\dagger} \partial_{\tau} \Psi - k^2 \Psi^{\dagger} \Psi  \biggr] +  \Psi^{\dagger} (\vec{b}\cdot\vec{\sigma}) \Psi +  \partial_{\tau} \Psi^{\dagger} (\vec{r}\cdot\vec{\sigma}) \partial_{\tau} \Psi  \biggr\}.
\label{two}
\end{equation}
In the non-interacting limit (i.e. when all the vectors are vanishing identically) the actions (\ref{twoa})--(\ref{two}) both coincide with the result obtained in Ref. \cite{fordp} and the two polarizations evolve independently. The physical model and the form of the interaction is specified by the components of the vectors $\vec{b}(k,\tau)$ and $\vec{r}(k,\tau)$. 

When Eq. (\ref{two}) involves the only diagonal Pauli matrix (i.e. $\sigma_{3}$) the action follows  from Eq. (\ref{two}) by choosing $\vec{b} = (0,\,0,\,b_{3})$ and 
$\vec{r} = (0,\,0,\,r_{3})$
\begin{eqnarray}
S_{pol} &=& \frac{1}{2} \int d^{3}k \int d\tau \biggl\{ a^2(\tau) \biggl[ \biggl( \partial_{\tau} h_{\otimes} \partial_{\tau} h_{\otimes}^{*} + \partial_{\tau} h_{\oplus} \partial_{\tau} h_{\oplus}^{*}\biggr) - k^2 \biggl(  h_{\otimes} h_{\otimes}^{*} +  h_{\oplus} h_{\oplus}^{*}\biggr)\biggr]  
\nonumber\\
&+& b_{3}(k,\tau) \biggl( h_{\oplus}^{*} h_{\oplus} - h_{\otimes} h_{\otimes}^{*}\biggr)
+ r_{3}(k,\tau) \biggl( \partial_{\tau} h_{\oplus}^{*}  \partial_{\tau}h_{\oplus} -  \partial_{\tau} h_{\otimes} 
\partial_{\tau} h_{\otimes}^{*}\biggr)\biggr\}.
\label{three}
\end{eqnarray} 
Depending upon the specific forms of $r_{3}$ and $b_{3}$ the evolution of $h_{\oplus}$ and $h_{\otimes}$  can 
be different,  but the corresponding equations for $h_{\oplus}$ and $h_{\otimes}$ do not mix. Conversely whenever 
the interaction involves either $\sigma_{2}$ or $\sigma_{1}$  the two polarizations are coupled in the linear basis. For instance  when  $\vec{b} = (0,\,b_{2},\,0)$ and $\vec{r} = (0,\,r_{2},\,0)$ Eq. (\ref{two}) becomes:
\begin{eqnarray}
S_{pol} &=& \frac{1}{2} \int d^{3}k \int d\tau \biggl\{ a^2(\tau) \biggl[ \biggl( \partial_{\tau} h_{\otimes}^{*} \partial_{\tau} h_{\otimes} + \partial_{\tau} h_{\oplus}^{*} \partial_{\tau} h_{\oplus}\biggr) - k^2 \biggl(  h_{\otimes}^{*} h_{\otimes} +  h_{\oplus}^{*} h_{\oplus}\biggr)\biggr]  
\nonumber\\
&+& i b_{2}(k,\tau) \biggl( h_{\oplus} h_{\otimes}^{*} - h_{\otimes} h_{\oplus}^{*}\biggr)
+ i r_{2}(k,\tau) \biggl( \partial_{\tau} h_{\oplus} \partial_{\tau} h_{\otimes}^{*} - \partial_{\tau} h_{\otimes} \partial_{\tau} h_{\oplus}^{*}\biggr)\biggr\}.
\label{foura}
\end{eqnarray} 
The action (\ref{foura}) becomes diagonal in the circular basis\footnote{The right (i.e. $R$) and left (i.e. $L$) polarizations are defined as 
$ e^{(R)}_{ij} = (e_{ij}^{\oplus} + i\,e_{ij}^{\otimes})/\sqrt{2}$  and $e^{(L)}_{ij} = (e_{ij}^{\oplus} - i\,e_{ij}^{\otimes})/\sqrt{2}$.  The relation between the linear and the circular tensor amplitudes 
follows easily from Eq. (\ref{one}).} where the Fourier amplitude of Eq. (\ref{one}) reads now $ h_{ij}(\vec{k},\tau) = [e^{(R)}_{ij} h_{R}(\vec{k},\tau) + e^{(L)}_{ij} h_{L}(\vec{k},\tau)]$, and 
$ h_{L} = (h_{\oplus} + i\, h_{\otimes})/\sqrt{2}$ and $h_{R} =  (h_{\oplus} - i\, h_{\otimes})/\sqrt{2}$. In the circular basis the action of Eq. (\ref{foura}) is:
\begin{eqnarray}
S_{pol} &=&  \frac{1}{2} \int d^{3}k \int d\tau \biggl\{ a^2(\tau) \biggl[  \biggl( \partial_{\tau} h_{R}^{*} \partial_{\tau} h_{R} + \partial_{\tau} h_{L}^{*} \partial_{\tau} h_{L}\biggr) - k^2 \biggl(  h_{R}^{*} h_{R} +  h_{L}^{*} h_{L}\biggr)\biggr] 
\nonumber\\
&+& b_{2}(k,\tau) \biggl( h^{*}_{R} h_{R} - h_{L}^{*} h_{L} \biggr)  + r_{2}(k,\tau)  \biggl( \partial_{\tau} h^{*}_{R} \partial_{\tau} h_{R} - \partial_{\tau} h_{L}^{*} \partial_{\tau} h_{L} \biggr)\biggr\}.
\label{four}
\end{eqnarray}
Once more, the two circular amplitudes will obey two different equations that  are however decoupled  and will eventually produce a net degree of polarization, as we shall more concretely illustrate in a moment. Needless to say that Eq. (\ref{four}) can be swiftly derived by working directly with bispinors; indeed  the action (\ref{two}) in the case $\vec{\sigma}= (0,\,\sigma_{2},\,0)$ is
\begin{equation}
S_{pol} = \frac{1}{2} \int d^{3} k \int d\tau \biggl\{ a^2(\tau)\biggl[\partial_{\tau} \Psi^{\dagger} \partial_{\tau} \Psi - k^2 \Psi^{\dagger} \Psi  \biggr] +  b_{2}(k,\tau) \Psi^{\dagger} \sigma_{2} \Psi +  r_{2}(k,\tau) \partial_{\tau} \Psi^{\dagger} \sigma_{2}  \partial_{\tau} \Psi  \biggr\},
\label{sixa}
\end{equation}
and it becomes diagonal by performing the following unitary transformation:
\begin{equation}
\Psi = U \Phi, \qquad U = \left(\matrix{1/\sqrt{2}
& 1/\sqrt{2}\cr i/\sqrt{2}&- i/\sqrt{2}}\right),  \qquad \Phi = \left(\matrix{h_{R}\cr h_{L}} \right).
\label{sixc}
\end{equation}
where $U^{\dagger} = U^{-1}$. Since $U^{\dagger} \sigma_{2} U = \sigma_{3}$ we  have that Eq. (\ref{two}) becomes:
\begin{equation}
S_{pol} = \frac{1}{2} \int d^{3} k \int d\tau \biggl\{ a^2(\tau)\biggl[\partial_{\tau} \Phi^{\dagger} \partial_{\tau} \Phi - k^2 \Phi^{\dagger} \Phi  \biggr] +  b_{2}(k,\tau) \Phi^{\dagger} \sigma_{3} \Phi +  r_{2}(k,\tau) \partial_{\tau} \Phi^{\dagger} \sigma_{3}  \partial_{\tau} \Phi  \biggr\}.
\label{sevena}
\end{equation}
Equation (\ref{sevena}) can be expressed in an even more compact form 
by introducing two appropriate matrices $Z$ and $W$:
\begin{equation}
S_{pol} = \frac{1}{2} \int d^{3}k \,\int d\tau\, \biggl[ \partial_{\tau} \,\Phi^{\dagger} \,\,Z\,\, \partial_{\tau} \,\Phi - \Phi^{\dagger}\,\, W \,\,\Phi \biggr].
\label{UN1}
\end{equation}
The two matrices appearing in Eq. (\ref{UN1}) are $Z(k,\tau)= \{[a^2(\tau) +  r_{2}(k,\tau) ] P_{R} +  [a^2(\tau) -  r_{2}(k,\tau) ] P_{L}\}$
and $W(k,\tau) = \{[k^2 a^2(\tau) -  b_{2}(k,\tau)] P_{R} +  [k^2 a^2(\tau) +  b_{2}(k,\tau)] P_{L}\}$
where $P_{L}= (I- \sigma_{3})/2$ and $P_{R} =  (I+ \sigma_{3})/2$ denote the left and right projectors while $I$ is the identity matrix. The same steps leading to Eqs. (\ref{sevena}) and (\ref{UN1}) can be repeated 
when the interaction is dictated by $\sigma_{1}$ rather than by $\sigma_{2}$. Since $\sigma_{1}$ has only real entries the analog of Eq. (\ref{sixa}) can be easily derived from Eq. (\ref{two}). The resulting action will only contain $r_{1}$ and $b_{1}$ and  can diagonalized by the following unitary transformation  
\begin{equation}
\Psi = V \Xi, \qquad V =  \left(\matrix{1/\sqrt{2}
& 1/\sqrt{2}\cr 1/\sqrt{2}&- 1/\sqrt{2}}\right), \qquad \Xi = \left(\matrix{h_{+}\cr h_{-}} \right),
\label{sixd}
\end{equation}
where  $\Xi$  is defined in the new basis provided by the sum and by the difference of the two linear polarizations (i.e.  $h_{\pm} = (h_{\oplus} \pm h_{\otimes})/\sqrt{2}$). By plugging Eq. (\ref{sixd}) into Eq. (\ref{two}) written in the case  $\vec{\sigma}= (\sigma_{1},\,0,\,0)$ we are now led to 
\begin{equation}
S_{pol} = \frac{1}{2} \int d^{3}k \, \int d\tau\, \biggl[ \partial_{\tau} \,\Xi^{\dagger}\,\, \widetilde{Z} \,\,\partial_{\tau} \,\Xi - \Xi^{\dagger} \,\,\widetilde{W} \,\,\Xi \biggr], 
\label{UN1a}
\end{equation}
where $\widetilde{Z}(k,\tau) = \{[a^2(\tau) +  r_{1}(k,\tau) ] P_{R} +  [a^2(\tau) -  r_{1}(k,\tau) ] P_{L}\}$, and $\widetilde{W}(k,\tau) = \{[k^2 a^2(\tau) -  b_{1}(k,\tau)] P_{R} +  [k^2 a^2(\tau) +  b_{1}(k,\tau)] P_{L}\}$ in full analogy with the results of Eq. (\ref{UN1}).  As in the case of the circular basis the components of $\Xi$ obey two different equations which are  decoupled.

While Eq. (\ref{two}) purportedly describes the most general interaction 
of the two tensor polarizations evolving in conformally flat 
backgrounds, the reverse must also be true and any concrete 
model will have to correspond to a specific choice of $\vec{b}(k,\tau)$ and $\vec{r}(k,\tau)$. 
Along this perspective the action of the relic gravitons may contain a parity-violating term \cite{par1,par2,par3} that involves the dual
the Riemann (or Weyl) tensor\footnote{A complementary class of examples is obtained by replacing the Riemann tensor with the Weyl tensor. Being the tracelss part of the Riemann tensor, the Weyl tensor vanishes for a spatially flat Friedmann-RobertsonÐWalker metric the derivation of the second-order action describing the tensor modes  is comparatively easier in the Weyl rather  than in the Riemann case that will be specifically studied hereunder.}: 
\begin{equation}
S = - \frac{1}{2 \ell_{P}^2} \int d^{4} x \, \sqrt{-g} \, R - \frac{\beta}{8} \int d^{4} x \sqrt{-g} f(\varphi) \widetilde{R}^{\mu\alpha\nu\beta} 
R_{\mu\alpha\nu\beta}, \qquad  \widetilde{R}^{\mu\alpha\nu\beta} = \frac{1}{2} E^{\mu\alpha\rho\sigma} R_{\rho\sigma}^{\,\,\,\,\,\,\,\,\nu\beta},
\label{nine}
\end{equation}
where $g$ is the determinant of the four-dimensional metric, $E^{\mu\alpha\rho\sigma} = \epsilon^{\mu\alpha\rho\sigma}/\sqrt{-g}$ and $\epsilon^{\mu\alpha\rho\sigma}$ is the Levi-Civita symbol;  $\beta$ is just a numerical constant while $f(\varphi)$ contains the dimensionless coupling to some scalar degree of freedom that can be identified, for instance,  with the inflaton or with some other spectator field.
The action of the tensor modes o the geometry follows, in this case, by perturbing 
Eq. (\ref{nine}) to second-order in the amplitude of the tensor modes of the geometry introduced prior to Eq. (\ref{one}). The explicit result is given by:
\begin{eqnarray}
\delta_{t}^{(2)} S &=& \frac{1}{8 \ell_{P}^2} \int d^{4}x \,\,a^2 \,\,\biggl[ \partial_{\tau}h_{ij}\,\,\partial_{\tau} h_{ij} - \partial_{k} h_{ij}\,\,\partial_{k}h_{ij} \biggr] 
\nonumber\\
&-& \frac{\beta}{8} \int d^{4} x \,\,(\partial_{\tau} f)\,\epsilon^{i\, j\,k}\,\, \biggl[ \partial_{\tau} h_{q i} \,\,\partial_{\tau} \partial_{j} h_{k q}  - \partial_{\ell} h_{i q} \,\,\partial_{\ell} \partial_{j} h_{q k} \biggr], 
\label{ten}
\end{eqnarray}
where $\epsilon^{i\,j\,k}$ is now the Levi-Civita symbol in three-dimensions; we remind that 
the Latin indices are all Euclidian. According to Eq. (\ref{one}) the tensor amplitudes appearing in Eq. (\ref{ten}) can be expressed in the linear polarization basis and the result willbe:
\begin{eqnarray}
\delta_{t}^{(2)} S&=& \frac{1}{2} \int d^{3} k \int d\tau \biggl\{ a^2 \biggl[ \partial_{\tau} h_{\oplus} \partial_{\tau} h_{\oplus}^{*} 
+ \partial_{\tau} h_{\otimes} \partial_{\tau} h_{\otimes}^{*} - k^2 \biggl( h_{\oplus} h_{\oplus}^{*} 
+  h_{\otimes} h_{\otimes}^{*} \biggr)\biggr]
\nonumber\\
&+& i k \beta \,\ell_{P}^2\, \partial_{\tau} f \biggl[  \partial_{\tau} h_{\oplus} \partial_{\tau} h_{\otimes}^{*} - \partial_{\tau} h_{\otimes} \partial_{\tau} h_{\oplus}^{*} - k^2 \biggl(  h_{\oplus} h_{\otimes}^{*}  -h_{\otimes} h_{\oplus}^{*}\biggr) \biggr] \biggr\}.
\label{eleven}
\end{eqnarray}
If the two linear polarizations appearing in Eq. (\ref{eleven}) are arranged into the components of the bispinor $\Psi$ we obtain a particular case\footnote{Indeed Eq. (\ref{sixa}) reproduces exactly Eq. (\ref{eleven}) provided $b_{2} = - k^3 \beta \ell_{P}^2 (\partial_{\tau} f)$ and $r_{2} =  k \beta \ell_{P}^2 (\partial_{\tau} f)$.} of Eqs. (\ref{two}) and (\ref{sixa}).  As argued in general terms in Eq. (\ref{four}), the resulting action becomes diagonal in the circular polarization basis and Eq. (\ref{eleven}) shall then be expressible 
in the compact form of Eq. (\ref{UN1}) where now the matrices $Z$ and $W$ are 
 given by $Z(k,\tau)  = \{[ a^2(\tau)  +  k \beta \, \partial_{\tau} f \,\ell_{P}^2]\,P_{R} + [ a^2(\tau)  -  k \beta \, \partial_{\tau} f \,\ell_{P}^2 ]P_{L}\}$ and by $W= k^2 Z$.

A different class of illustrative models is obtained by considering  the case where only $\vec{b}$ does not vanish. While these examples would seem naively difficult to concoct, they may arise from  the following generally covariant action: 
\begin{equation}
S = - \frac{1}{2 \ell_{P}^2} \int d^{4} x \sqrt{-g} \, R - \frac{1}{2\ell_{P}^2 M^4} \int d^4 x f(\varphi) R_{\mu\alpha\nu\beta} \, Y^{\mu\alpha} \, \widetilde{Y}^{\nu\beta}, 
\label{fifteenth}
\end{equation}
where $Y_{\mu\nu}$ and  $\widetilde{Y}^{\mu\nu} = E^{\mu\nu\rho\sigma} \, Y_{\rho\sigma}/2$ are the gauge field strength and its dual. In Eq. (\ref{fifteenth})  $M$ sets the typical scale of the interaction. Note, incidentally, that the explicit powers of $M$ are often omitted in the analysis of effective theories of inflation \cite{par1} with the proviso that all constants in the higher derivative terms of the effective action take values that are powers of $M$ indicated by dimensional analysis, with coefficients roughly of order unity.  If there exist a family of four-dimensional observers moving with four-velocity $u^{\mu}$
(possibly related with the covariant gradients of a scalar field) in Eq. (\ref{fifteenth}) the gauge fields can be covariantly decomposed in their electric and magnetic 
parts according to $Y_{\mu\alpha} = {\mathcal E}_{[\mu} u_{\alpha]} + 
E_{\mu\alpha\rho\sigma} u^{\rho} {\mathcal B}^{\sigma}$ and to $
\widetilde{Y}^{\nu\beta} = {\mathcal B}^{[\nu} u^{\beta]} + E^{\nu\beta\rho\sigma} {\mathcal E}_{\rho} u_{\sigma}$ (note that $[\,.\,.\,.\,]$ denotes an antisymmetric combination of the two corresponding tensorial indices).
Inserting this decomposition into Eq. (\ref{fifteenth}) and neglecting the electric contributions\footnote{To zeroth 
order in the tensor amplitude we have $\widetilde{Y}^{0\,i} = - b^{i}/a^2$ and $Y^{ij} = - \epsilon^{i\,j\,k} b_{k}/a^2$. To first and second-order in the amplitude of the tensor modes the previous expressions 
are corrected as $\delta_{t}^{(1)} \widetilde{Y}^{0i} = h^{i k} b_{k}/a^2$ and as $\delta_{t}^{(2)} \widetilde{Y}^{0i} =- h^{i \ell} h_{\ell}^{\,\, k} b_{k}/a^2$. }
we can perturb the action to second-order in the amplitude of the tensor modes of the geometry and the result of this step can be written as:
\begin{eqnarray}
\delta_{t}^{(2)} S &=& \frac{1}{8\ell_{P}^2} \int d^{4} x \biggl\{ a^2 \biggl[ (\partial_{\tau} h_{ij}) (\partial_{\tau} h_{ij})- (\partial_{k} h_{ij}) (\partial_{k} h_{ij}) \biggr] 
\nonumber\\
&+& 4 a^2 F \,n_{c} \,n_{a} \,\epsilon^{b p c}\biggl[ 
\partial_{\tau} h_{p q} ( \partial_{a} h_{q b} - \partial_{b} h_{q a} ) -  \partial_{\tau} h_{a q}
(\partial_{b} h_{q p} - \partial_{p} h_{b q})\biggr] \biggr\},
\label{SPN1}
\end{eqnarray}
where $b_{a}(\tau) = n_{a} b(\tau)$ and  $F= (b^2 \, f)/M^4$. The action perturbed to second-order in the amplitude of the tensor modes of the geometry becomes then
\begin{eqnarray}
S_{pol} &=& \frac{1}{2} \int d^{3} k \int d\tau \biggl\{ a^2 \biggl[ \biggl( \partial_{\tau} h_{\oplus} \partial_{\tau} h^{*}_{\oplus} + \partial_{\tau} h_{\otimes} \partial_{\tau} h^{*}_{\otimes} \biggr) - k^2 
\biggl( h_{\oplus} h^{*}_{\oplus} + h_{\otimes} h^{*}_{\otimes} \biggr)\biggr] 
\nonumber\\
&+& i k a^2 F \biggl[ (\partial_{\tau} h_{\oplus}) h_{\otimes}^{*} -   (\partial_{\tau} h_{\otimes}) h_{\oplus}^{*}\biggr]
+ i k a^2 F \biggl[ h_{\oplus} (\partial_{\tau}  h_{\otimes}^{*}) -   h_{\otimes} (\partial_{\tau} h_{\oplus}^{*})\biggr]\biggl\}.
\label{SPN3}
\end{eqnarray}
In the linear basis Eq. (\ref{SPN3}) coincides with Eq. (\ref{twoa}) having chosen $\vec{p}(k,\tau) = (0,\,\, k a^2 F,\,\,0)$. 
Up to a total derivative the obtained equation can be brought in the form (\ref{two}) with $\vec{b}(k,\tau) =[0,\, - k \partial_{\tau}(a^2 F),\,0]$ and then diagonalized in the circular basis. If the two previous steps are inverted the final result does not change and Eq. (\ref{SPN3}) can be diagonalized\footnote{According to Eq. (\ref{sixc}) we can go from the linear to the circular basis by positing $\Psi = U \, \Phi$ where $\Phi$ denotes the bispinor in the circular basis. Since $U^{\dagger} \, \sigma_{2} U = \sigma_{3}$
the canonical action is easily obtained.} before the elimination of the total derivative. In either case the final form 
of the action (\ref{SPN3}) is 
\begin{eqnarray}
S_{pol}  =\frac{1}{2} \int d^{3} k \int d\tau \biggl\{ a^2 \biggl[ \partial_{\tau} \Phi^{\dagger} \partial_{\tau} \Phi - k^2 \Phi^{\dagger} \Phi \biggr] -  k \partial_{\tau} (a^2 F)  \Phi^{\dagger} \sigma_{3} \Phi \biggr\}.
\label{SPN6}
\end{eqnarray}
Equation (\ref{SPN6}) can be finally put in the form (\ref{UN1}) by choosing 
$Z = I\, a^2$ and  $W = \{[k^2\, a^2+ k\,\partial_{\tau} (a^2 F)] P_{R} + [k^2\,a^2 - k\,\partial_{\tau} (a^2 F)] P_{L}\}$.

In the two previous paragraphs different generally covariant models have been shown to reproduce some particular cases of Eq. (\ref{two}) and this conclusion corroborates the validity of the direct derivation.  We now turn to the evaluation of the degree of polarization and, for this purpose, it is interesting to remark that  Eq. (\ref{UN1})  can be simplified even further by defining the rescaled bispinor ${\mathcal M} = (P_{R} z_{R} + P_{L} z_{L}) \Phi $ whose components, in the circular basis, are given,  by $\mu_{R} = z_{R} h_{R}$ and $\mu_{L} = z_{L} h_{L}$. After expressing Eq.  (\ref{UN1}) in terms of ${\mathcal M}$ we obtain: 
\begin{equation}
S_{pol} = \frac{1}{2} \int d^{3} k \int d\tau \biggl\{ \partial_{\tau} {\mathcal M}^{\dagger} \partial_{\tau}{\mathcal M} + 
{\mathcal M}^{\dagger} \biggl[ \biggl( \frac{z_{R}^{\prime\prime}}{z_{R}} - \omega_{R}^2 \biggr) P_{R} + 
 \biggl( \frac{z_{L}^{\prime\prime}}{z_{L}} - \omega_{L}^2 \biggr) P_{L} \biggr] {\mathcal M} \biggr\},
 \label{SPN8}
 \end{equation}
where the prime denotes a derivation with respect to the conformal time coordinate and the same shorthand notation will be employed hereunder. Furthermore, in Eq. (\ref{SPN8}) $z_{R,\,\,L}$ and $\omega_{L,\,\,R}$ are defined as:
\begin{eqnarray}
z_{R}^2(k,\tau) &=& a^2(\tau) +  r_{2}(k,\tau), \qquad z_{L}^2(k,\tau) = a^2(\tau) -  r_{2}(k,\tau),
\nonumber\\
\omega_{R}^2(k,\tau) &=& \frac{k^2 a^2(\tau) - b_{2}(k,\tau)}{a^2(\tau) + r_{2}(k,\tau)}, \qquad 
\omega_{L}^2(k,\tau) = \frac{k^2 a^2(\tau) + b_{2}(k,\tau)}{a^2(\tau) - r_{2}(k,\tau)}.
\label{SPN10}
\end{eqnarray}
For example the action of Eq. (\ref{nine}) implies 
that $\omega_{L} = \omega_{R} = k^2$ while in the case of Eq. (\ref{fifteenth}) $r_{2} \to 0$ and 
$z_{L} = z_{R} = a (\tau)$. Similarly the general expressions of Eq. (\ref{SPN10}) may simplify for other specific values of $b_{2}(k,\tau)$ and $r_{2}(k,\tau)$. Recalling that $\Phi$ denotes the bispinor in the circular basis the degree of  circular polarization can be defined as:
\begin{equation}
\Pi_{circ}(k,\tau) = \frac{\Phi^{\dagger} \, \sigma_{3} \, \Phi}{\Phi^{\dagger}\, \Phi}  = \frac{| h_{R}(k,\tau)|^2 - |h_{L}(k,\tau)|^2}{| h_{R}(k,\tau)|^2 + | h_{L}(k,\tau)|^2}.
\label{DP2}
\end{equation}
From Eq. (\ref{SPN8}) the evolution of ${\mathcal M}$ reads
\begin{equation}
\partial_{\tau}^2 {\mathcal M} + \biggl[ \biggl(\omega_{R}^2 - \frac{z_{R}^{\prime\prime}}{z_{R}} \biggr) P_{R} + 
 \biggl(\omega_{L}^2- \frac{z_{L}^{\prime\prime}}{z_{L}}  \biggr) P_{L} \biggr] {\mathcal M} =0.
 \label{DP2a}
 \end{equation}
Equation (\ref{DP2a}) reduces  to a pair of decoupled equations defined in the circular basis: 
\begin{equation}
\mu_{R}^{\prime\prime} + \biggl[ \omega_{R}^2 - \frac{z_{R}^{\prime\prime}}{z_{R}}\biggr] \mu_{R}=0, \qquad 
\mu_{L}^{\prime\prime} + \biggl[ \omega_{L}^2 - \frac{z_{L}^{\prime\prime}}{z_{L}}\biggr] \mu_{L}=0,
\label{DP2b}
\end{equation}
where we defined $\mu_{R} = z_{R}\,h_{R}$ and $\mu_{L} = z_{L}\, h_{L}$.
The evolution of the right and of the left movers can be studied within different approximation schemes.  The expansion in the conformal coupling parameter, originally explored by Birrell  and Davies \cite{birrell}, has been subsequently applied to the case of gravitational waves by Ford \cite{ford2}.   However, since the illustrative goal is to evaluate the degree of circular polarization at high-frequencies, the WKB approximation  \cite{gr2} seems more directly applicable. While this method has not been applied so far to the polarized case,  this gap will now be bridged, at least partially. The equations for $\mu_{R}$ and $\mu_{L}$ reported in (\ref{DP2b}) closely resemble  Schr\"odinger equations with different $k$-dependent potentials  for the left and right movers [i.e.  $V_{R}(k,\tau) = z_{R}^{\prime\prime}/z_{R}$  and $V_{L}(k,\tau) = z_{L}^{\prime\prime}/z_{L}$]. Thus for $\omega_{R}^2 \gg V_{R}$ and $\omega_{L}^2 \gg V_{L}$ the general solution of Eq. (\ref{DP2b}) is\footnote{In general terms since $b_{2}/r_{2} = - k^2$,  we shall also have that $\omega_{L}^2 = \omega_{R}^2 = k^2$.}
\begin{equation}
\mu_{X}(k,\tau) = \frac{1}{\sqrt{2 \omega_{X}}} \biggl[ \alpha_{X} e^{- i \int^{\tau} \omega_{X}(k,\tau^{\prime}) d \tau^{\prime}} + \beta_{X} e^{ i \int^{\tau} \omega_{X}(k,\tau^{\prime}) d \tau^{\prime}}\biggr], \qquad \omega_{X}^2 \gg \biggl| \frac{z_{X}^{\prime\prime}}{z_{X}} \biggr|,
\label{DP3}
\end{equation}
where $\alpha_{X}$ and $\beta_{X}$ are arbitrary complex numbers and
$X=R,\,\,L$; Eq. (\ref{DP3}) holds independently for the left and right 
movers  provided the variation of $\omega_{X}$ is sufficiently slow (i.e. $|V_{X}| \ll \omega_{X}^{\prime}/\omega_{X} < |z_{X}^{\prime}/z_{X}|$). In the opposite limit (i.e. $\omega_{X}^2 \ll V_{X}$) the general solution of Eq. (\ref{DP2b}) is instead given by:
\begin{equation}
\mu_{X}(k,\tau) = A_{X}(k) z_{X}(k,\tau) + B_{X}(k) z_{X}(k,\tau) \int^{\tau} \frac{d\tau^{\prime}}{ z^2_{X}(k,\tau^{\prime})}, \qquad 
\omega_{X}^2 \ll \biggl| \frac{z_{X}^{\prime\prime}}{z_{X}} \biggr|.
\label{DP4}
\end{equation}
For a wide class of problems the potentials $|V_{X}|$ have a bell-like shape in the conformal time coordinate and vanish in the limit $\tau \to \pm \infty$. The solution (\ref{DP3}) is valid outside the potential barrier $|V_{X}|$ (i.e. inside the effective horizon defined by the variation of $z_{X}$); the solution (\ref{DP4}) holds instead when $|V_{X}|$ dominates against $\omega_{X}^2$ (or, more precisely, when the corresponding wavelengths are larger than the effective horizon). The turning points are fixed by $\omega_{X}^2 = V_{X}$ and will be denoted by $\tau_{ex}$ (i.e. the time at which the mode {\em exits}) the effective horizon) and by $\tau_{re}$ (i.e. the moment at which the given mode {\em reenters} the effective horizon). To the left of the barrier, for $\tau < \tau_{ex}$ the solution will be in the form $e^{- i \int^{\tau} \omega_{X}(k,\tau^{\prime})d\tau^{\prime}}/\sqrt{2 \omega_{X}}$. To the right of the barrier the solution is instead given by Eq. (\ref{DP3}). Finally, between the turning points the solution has the form (\ref{DP4}). The continuous matching of the three solutions (and of their first derivatives) across the two turning points allows for an explicit determination of $\alpha_{X}$ and $\beta_{X}$. In the interesting physical case  (i.e. $|z_{re}^{(X)}/z_{ex}^{(X)}|\gg 1$) the coefficient $\beta_{X}$ is always larger than $\alpha_{X}$
(i.e. $|\beta_{X}|^2 \gg |\alpha_{X}|^2$); thus we shall only need to mention the results for $|\beta_{R}|^2$ and $|\beta_{L}|^2$:
\begin{eqnarray}
|\beta_{L}|^2 &\simeq& \frac{1}{4} \biggl[\frac{z^{(L)}_{re}}{z^{(L)}_{ex}}\biggr]^2 \biggl[1 + 
\frac{{\mathcal L}_{re}^2}{\omega_{L}^2} \biggr] 
\biggl\{ 1 - 2 {\mathcal L}_{ex} z^{(L)\,2}_{ex} {\mathcal J}_{L} 
+ z^{(L)\,\,4}_{ex} \biggl[ {\mathcal L}_{ex}^2 
+ \omega_{L}^2\biggr] {\mathcal J}_{L}^2\biggr\},
\nonumber\\
|\beta_{R}|^2 &\simeq& \frac{1}{4} \biggl[\frac{z^{(R)}_{re}}{z^{(R)}_{ex}}\biggr]^2 \biggl[1 + 
\frac{{\mathcal R}_{re}^2}{\omega_{R}^2} \biggr]\biggl\{ 1 - 2 {\mathcal R}_{ex} z^{(R)\,2}_{ex} {\mathcal J}_{R}
+ z^{(R)\,\,4}_{ex} \biggl[ {\mathcal R}_{ex}^2 + \omega_{R}^2\biggr] {\mathcal J}_{R}^2\biggr\}.
\label{DP6}
\end{eqnarray}
In Eq. (\ref{DP6})  the rates of variation of $z_{R}$ and $z_{L}$ (i.e. ${\mathcal R} = z_{R}^{\prime}/z_{R}$ and ${\mathcal L} = z_{L}^{\prime}/z_{L}$) have been introduced while ${\mathcal J}_{L} (k,\tau_{ex},\tau_{re})$ and ${\mathcal J}_{R} (k,\tau_{ex},\tau_{re})$ involve two different integrals between the two turning points:
\begin{equation}
 {\mathcal J}_{L}(k,\,\tau_{ex}, \tau_{re}) = \int_{\tau_{ex}}^{\tau_{re}} \frac{d\tau^{\prime}}{z_{L}^2(k,\tau^{\prime})}, \qquad {\mathcal J}_{R}(k,\tau_{ex}, \tau_{re}) = \int_{\tau_{ex}}^{\tau_{re}} \frac{d\tau^{\prime}}{z_{R}^2(k,\tau^{\prime})}.
\label{DP7}
\end{equation}
If the interaction between the polarizations ceases after the end of inflation, 
the high frequencies will cross the barrier the second time when
$z_{R} = z_{L} = a(\tau)$ and will remain inside the effective horizon thereafter. In this 
case  the total degree of circular polarization of Eq. (\ref{DP2}) can be expressed as: 
\begin{equation}
\Pi_{circ} =\frac{ |\beta_{R}(k)|^2 - |\beta_{L}(k)|^2 }{|\beta_{R}(k)|^2 + |\beta_{L}(k)|^2}.
\label{DP8}
\end{equation}
Equation (\ref{DP8}) demonstrate that the left and right movers see effectively different potential barriers so that the degree 
of circular polarization is ultimately determined by the difference of the two dominant mixing coefficients. The explicit evaluation of Eqs. (\ref{DP2}) and (\ref{DP8}) is delicate when the potential $V_{X}\to 0$ in the vicinity of the second turning point (as it happens when the second crossing takes place during radiation where $a^{\prime\prime}=0$ \cite{gr2}). With these caveats  after inserting Eq. (\ref{DP6})  into Eq. (\ref{DP8}) the explicit expression of the polarization degree
depends solely on the values of the pump fields and of their first derivatives at the turning points:
\begin{equation}
\Pi_{circ} = \frac{|z_{re}^{(R)}|^2\, |z_{ex}^{(L)}|^2 (k^2 + {\mathcal R}_{re}^2) - |z_{ex}^{(R)}|^2\, |z_{re}^{(L)}|^2 (k^2 + {\mathcal L}_{re}^2) }{|z_{re}^{(R)}|^2 |z_{ex}^{(L)}|^2 (k^2 + {\mathcal R}_{re}^2) + |z_{ex}^{(R)}|^2 |z_{re}^{(L)}|^2 (k^2 + {\mathcal L}_{re}^2)}.
\label{DP9}
\end{equation}
If all the modes reenter after the end of inflation (i.e. $z_{re}^{(R)} = z_{re}^{(L)} = a_{re}$) polarization degree is particularly 
simple and it only depends on $z_{ex}^{(X)}$, i.e.  $\Pi_{circ} =(|z_{ex}^{(L)}|^2- |z_{ex}^{(R)}|^2 )/(|z_{ex}^{(L)}|^2+ |z_{ex}^{(R)}|^2)$.  

While Eqs. (\ref{DP8}) and (\ref{DP9}) hold for different functional forms of the pump fields, more explicit expressions also demand further details on  $z_{L}$ and $z_{R}$. If we consider, for the sake of illustration, the case  $z_{R} = \sqrt{a^2 + k \beta f^{\prime} \ell_{P}^2}$ and $z_{L} = \sqrt{a^2 - k \beta f^{\prime} \ell_{P}^2}$ (discussed in Eq. (\ref{eleven}) and thereunder) the explicit form of Eq. (\ref{DP9}) becomes\footnote{Note that ${\mathcal A}_{s}$ denotes the amplitude of the scalar power spectrum appearing since $(H/M_{P}) = \sqrt{\pi \epsilon {\mathcal A}_{s}}$.} 
\begin{equation}
 |\Pi_{circ}| =  \beta  \sqrt{2 \epsilon} \biggl(\frac{H}{\overline{M}_{P}}\biggr)^2 
 = 6 \times 10^{-13} \biggl(\frac{\beta}{0.1}\biggr) \biggl(\frac{\epsilon}{0.001}\biggr)^{3/2} \biggl(\frac{{\mathcal A}_{s}}{2.41\times 10^{-9} }\biggr),
 \label{DP11}
 \end{equation}
 where we assumed, for illustration, that the dependence on the inflaton of $f(\varphi)$ is linear, (i.e. $f = \ell_{P} \varphi$) and that   $z_{re}^{(R)} = z_{re}^{(L)} = a_{re}$. In Eq. (\ref{DP11}) $\epsilon$ is the slow-roll parameter that ultimately determines the derivative of $\varphi$ (i.e. $\varphi^{\prime} = \sqrt{ 2 \epsilon} \, a \, H \overline{M}_{P}$).  The result of Eq. (\ref{DP11}) becomes 
a bit different if $z_{re}^{(R)} \neq z_{re}^{(L)} \neq a_{re}$. In this case Eq. (\ref{DP9}) implies
\begin{equation}
|\Pi_{circ}| = \beta \biggl(\frac{H}{\overline{M}_{P}}\biggr)^2 \biggl| \sqrt{2 \epsilon} - \sqrt{3 w_{re} +1} \biggl(\frac{k}{k_{1}}\biggr)^{ \frac{3 (w_{re} +1)}{ 3 w_{re} +1}}\biggr|,
\label{DP13}
\end{equation}
where $w_{re}$ denotes the barotropic index of the plasma when the given mode reenters the effective horizon (i.e. at the second turning point). While Eqs. (\ref{DP11}) and  (\ref{DP13}) are qualitatively different, they are similar from the quantitative and physical viewpoints.  They both hold for all the modes that reentered the effective horizon after the end of inflation but before the onset of the matter epoch (i.e. for frequencies larger than $100$ aHz). In practice, however, the estimates apply for  frequencies  larger than the Hz since we also neglected the presence of the neutrino anisotropic stress \cite{anis1}. They also suggest that $\Pi_{circ}$ gets larger at high frequencies. More specifically since  $k_{1}$ is of the order of the GHz (and it corresponds to the Hubble rate at the end of inflation) the degree of polarization is maximal at high frequencies so that the instruments  operating in the kHz (or even MHz) regions are potentially more promising than the space-borne interferometers operating below the Hz. 

All in all the polarizations of the tensor modes  evolving in cosmological backgrounds have been described in terms of appropriate bispinors closely analog to what are commonly referred to as Jones vectors. Unlike the case of polarized optics, the present goal was to obtain a gauge-invariant action containing at most two (conformal) time derivatives and reducing to the standard (Ford-Parker) result when the two polarizations are mutually decoupled and only feel the overall effect of the space-time curvature. After arguing that the interactions potentially leading to polarized relic gravitons can be compactly classified in general terms, we showed that the reverse is also true. For this purpose the direct derivation  has been corroborated by few examples demonstrating, in a conservative perspective, that different classes of generally covariant models ultimately fit within the scheme of the spinor action proposed here.  As an illustrative application we derived the degree of polarization and its spectral dependence at high frequencies by introducing a suitable WKB approximation where the modes corresponding to each of the two tensor amplitudes obey a different evolution equation and cross their effective horizons at slightly different turning points.

The author wishes to thank T. Basaglia and A. Gentil-Beccot of the CERN Scientific Information Service for their kind assistance.

\end{document}